\documentclass[11pt,dvipdfm]{article}
\usepackage{times,graphicx}
\usepackage{url}
\title{Unstructured and structured data:\\Can we have the best of both worlds with\\ large language models?}
\author{Wang-Chiew Tan\footnote{The opinions presented are my own.}\\Reality Labs Research, Meta\\{\tt wangchiew@meta.com}}

\date{}

\begin{document}

\maketitle

\section{Introduction}
We are witnessing rapid advancements in the area of large language models (LLMs). A search on Google Scholar shows there are about 3,910 papers with ``large language models'' in the titles of the papers in 2022. As of April 12, 2023, there are already 1700 articles with LLMs in their titles. In addition to Google Scholar, we are also witnessing huge volumes of blog posts, news articles, twitter feeds, and open-source repositories around LLMs that have sprung up in recent months. 


Perhaps ChatGPT (released on November 30, 2022) is the epitome of this LLM revolution that truly unleashed and showcased, to the masses, the power of what has been brewing in the natural language and machine learning communities in recent years. There are many things ChatGPT\footnote{I use the term ``ChatGPT'' as a representative for general chat systems based on LLMs for the rest of this article.} can do and does so impressively by generating intelligent human-like responses to your questions. Through natural language as input and output\footnote{GPT-4 can accept image and text inputs.}, it can solve non-trival mathematical problems, to a certain extent~\cite{frieder2023mathematical,shakarian2023independent}, translate your specification into code in a programming language of your choice (e.g.,~\cite{kashefi2023chatgpt,shakarian2023independent}), help you write proses in different styles and the list of accolades goes on~\cite{nyt-chatgptuses}.  

In addition to its ability to answer questions, one can also prompt it with tables, such as CSV files, and ask questions over the tables in natural language. More interestingly, it is also possible to prompt ChatGPT with both text and tables and ask questions over the two types of information seamlessly. A little more perseverance in this exercise quickly reveals that ChatGPT has a limit on how much one can input with each prompt and how much information it will retain, at least based on my experience when I tried this at the end of March 2023.

It is natural to wonder whether we can use ChatGPT, with some extensions, as a system for storing and querying data, with natural language as the primary medium of input and output, which it excels at. What are the challenges of doing so, and how can we, as database practitioners and theoreticians, make progress in this context?

\section{Unstructured and structured data}
A lot of data, including text, images, audio, and videos, sits ``outside the box'' today. Often, such unstructured data contain multiple modalities simultaneously. For example, we often find text in images~\cite{DBLP:conf/nips/KielaFMGSRT20}, and we may also find text associated with videos and/or images on the web. Unstructured data is prevalent to a large extent because it is easily authored and shared by users~\cite{HalevyCIDR03} through a variety of apps and authoring tools that are widely available.
Such data is often queried with keyword search and today, they can also be queried in natural language with LLMs.
Typically the cost of devising a schema and setting up a database for querying the data inhibits the use of a database management system (DBMS) upfront. However, as data scales, the need for structure and semantics becomes more critical, so as to enable faster and more accurate retrieval of content. 
For example, organizing photos by year, trips, or entity types (such as people or pets) adds some structure, which makes answering certain types of queries much faster and more accurate. However, answering complex queries such as ``{\it when was the last time I went to the coffee shop beside restaurant Italio?}'' or ``{\it how many times did I celebrate Anna's birthday with a mango cake?}'' requires non-trivial reasoning and computation over the data that goes beyond the capabilities of LLMs today~\cite{timelineqa, neuraldb}.

Enterprise data sits on the other end of the spectrum. It is, for the most part, not authored by everyday users and is highly structured, often sitting "in a box" in some DBMS. Enterprise data comes with a well-devised schema and is typically highly optimized to serve a sizable query workload with great efficiency.
Such data is often queried directly with SQL which is adequately expressive for specifying complex queries such as those with aggregates and recursion. Significant research has also been carried out to enable querying a database with natural language (e.g., ~\cite{qi-etal2022rasat, scholak-etal-2021-picard}), where questions are posed in natural language and translated into SQL, which can then be executed over the DBMS. A DBMS is highly optimized to handle large amounts of data and can also perform transactions with ACID guarantees~\cite{acid}. However, the core dbms does not understand natural language and it tends to fall short in its ability to query unstructured data, which are often stored as blobs and adding semantics to the blobs require additional effort. For example, ``{\it show me the sales numbers over Black Friday and Cyber Monday last year}'' requires commonsense knowledge on what ``sales numbers'' mean and how that maps to relevant table attribute(s), which may be stored under different names in different databases, and when Black Friday and Cyber Monday occurred. Another example is "{\it find all items with good reviews that are similar to these images}," which requires matching images (semantically) and interpreting what ``good reviews'' mean based on the data.

Despite the divide in how unstructured and structured data are managed today, the desire to query in natural language is common to both. At the same time, it is unreasonable and unnatural to expect all unstructured data to fit in some structure, or vice versa to leverage one system for querying. So, how can we effectively query both types of data with the help of LLMs, which possess tremendous knowledge and language understanding? 

\section{The best of both worlds with LLMs?}

LLMs contain tremendous {\em parametric} knowledge in their model parameters but lack the ability to incorporate external data (i.e., data outside their model).   Hence, if a model is trained based on data up to, say, 2021, it will not provide correct answers about events or facts that require knowledge after 2021. For example, if someone asks the question "{\it How were the midterm election results of 2022?}" on the webpage with {\tt text-davinci-003}~\cite{openai-text-davinci-003}, the answer returned is "{\it The midterm election results of 2022 are not yet available, as the election has not yet taken place}." This is because {\tt text-davinci-003} is trained with data up to June 2021.

Retrieval-augmented language models (e.g., ~\cite{guu-etal-realm, izacard_few-shot_2022, yasunaga2022retrievalaugmented}) overcome this limitation by adopting a {\em semi-parametric} approach to answering queries. They use external data, by first
retrieving relevant data from an external data store, 
and then attempts to answer a question conditioned on the retrieved data and with their parametric knowledge. However, as demonstrated by (Table 5,~\cite{timelineqa}), such systems can still perform poorly on complex queries involving aggregates and certain types of temporal queries. This is because, for such queries, oftentimes, it is impossible to fit all necessary data for answering the question into the finite-length token input imposed by language models. 
However, as LLMs get even larger or as more advances are made to increase the token limits imposed by language models, one can anticipate that LLMs will take larger and larger inputs in future and the finite-length token limit may no longer be an issue soon. At the same time, it is likely that there will be even larger datasets to manage and an even larger set of data is required for computing the right answers. So this problem will persist, at least for a while.

A proposal to overcome the above limitation for some types of queries is described in~\cite{posttext}. The paper presents a vision of using views (e.g., as tables) to structure portions of the underlying data sources. The data sources may be of different modalities, such as text, images, videos, or even tables. Views are used to surface important properties about data or associations between data of different modalities and LLMs are used to translate natural language queries into queries (e.g., SQL in this case) that can be executed over the views whenever possible. With this proposal, a key question is to understand when a natural language query can be answered with views. If a query cannot be answered using views, the system falls back to retrieval-augmented language models to answer the query to its best effort. Alternatively, the two components (view-based and retrieval-based query answering components) can also collaborate to produce a final answer. There are several other questions raised in the paper, such as what views should be materialized? How can one automatically select the ``right'' views to materialize given an anticipated query workload? And how can one decide when a question is better answered with views, the retrieval system, or even both?

In addition to the above, I will highlight below what I believe are some of the more pertinent questions that may be of immediate interest to the database community:

\smallskip
\noindent
{\bf Query answering with different resources and budget constraints~} The topic of finding a good query plan to answer an given query was already discussed in ~\cite{posttext}. The core of that system relies on two components --- views and a retrieval-augmented language model for answering queries. It is conceivable to augment the system with additional components, such as one that generates images given a natural language prompt, which may sometimes be useful for answering certain queries\footnote{As they say, a picture is worth a thousand words.}. 
A key question then becomes how do we understand when to leverage which component for answering a query or have the components collaborate to derive an answer? Furthermore, LLMs are compute-intensive and can be slow in generating an answer. Some LLMs are also not free. In addition to deciding how best to answer a query with all the available resources, how can one account for the strengths and limitations of each component to enumerate and compare plans for computing an answer to a given query and/or under a given budget? Can we also use a language model to generate a query plan or some parts of it, similar to how language models have been used to self-reason a sequence of steps to derive answers from questions (e.g.,~\cite{wei2022chain, yao2023react})?

\smallskip
\noindent
{\bf Provenance~}
Provenance is well-studied for certain classes of SQL queries, and is roughly defined as the source tuples that explains why a tuple is in the result of the query. As mentioned in~\cite{posttext}, one should attempt to answer queries using table views whenever possible by translating the natural language query into a SQL query that can be executed over the views. This way, it is possible that provenance can be obtained ``for free''. However, SQL queries that are generated by LLMs can be complex, for example, with nested SQL queries and/or aggregates in the FROM or WHERE clauses. For such cases, can we decompose the generated SQL query into a sequence of one or more ``simpler queries'' instead, where the provenance for simpler queries is well-understood and can be derived easily? If this is not possible, can we strategize a plan for answering the query in a different way so that provenance can be derived? 
The problem of finding an alternative query plan is related to the discussion in the earlier paragraph, but here the focus is on deriving a plan with sufficiently simple steps to enable provenance.

Retrieval-augmented language models also provide more guarantees for providing evidence for their answers. We are also beginning to witness implementations where sources of answers presented by retrieval-augmented language models are returned as part of the answers~\cite{retrievalqawithsources}. However, more research needs to be carried out to attribute provenance to training data (e.g., ~\cite{akyurek-etal-2022-towards, han-etal-2020-explaining, tracin2020}) to form a more comprehensive picture of provenance for the provided answer.


\smallskip
\noindent
{\bf Prompt Engineering Analysis~}
LLMs have limits on the number of tokens they can take as input. Even if one takes advantage of all the tokens one can use, prompting them with more information does not always translate to better answers as sometimes, presenting the language model with more information confuses the language model. This means one needs to be judicious in what we send to a language model for it to derive a correct answer of high quality\footnote{There are varied ways to answer a question correctly. Some are better than others.}. 

The answers returned by language models are also sensitive to how they are prompted. Sometimes the same question phrased slightly differently will result in completely different answers. In {\em prompt engineering}, the goal is to find the best prompt for the task at hand. 

Given the maximum token limit of language models, can we optimize the answers returned by a language model by strategically summarizing relevant data and/or removing irrelevant data? 
For example, the entity matching system of~\cite{Ditto}, which uses a language model, immediately performs entity matching more accurately when text descriptions of data are strategically shortened using a simple trick; by keeping only tokens of value, with high TD/IDF.
On a more theoretical side, given a suitable definition of what is a prompt and assumptions about language models, can we characterize what types of queries that require access to external data can be answered with one prompt, a finite number of prompts, or an asymptotic number of prompts under a budget of tokens?

\section{Conclusion}
The field of LLMs is moving fast, both in research and industry. In addition to~\cite{posttext}, \cite{halevy-yu} have also described the challenges of answering queries, in the context of augmented language models and data integration, with a single source or by chaining multiple sources. In~\cite{saeed2023querying}, the authors described how a DBMS can be extended to leverage LLMs to improve query answering and also pointed to the direction of a hybrid query answering system involving both a DBMS and LLMs. I believe this is only the beginning and we will see many more visions, research, and implementations soon in this area of query answering systems that embrace both unstructured and structured data through the use of large language models.

\medskip
\noindent
{\bf Acknowledgements~} Many of the ideas above are inspired from discussions with Alon Halevy and Yuliang Li. I also thank many of my colleagues at Meta --- Lambert Mathias, Richard Newcombe, and Luna Dong --- for active discussions around large language models from which I have learnt lots and also to Lucian Popa for his feedback on this article.

\small
\bibliographystyle{abbrv}
\bibliography{bib}

\end{document}